\documentstyle[12pt,psfig]{article}

\begin{document}

\newcommand{\rplus}{\rangle_{{}_+}}

\newcommand{\ECM}{\em Departament d'Estructura i Constituents de la
Mat\`eria\\ 
Universitat de Barcelona\\ Diagonal 647, E-08028 Barcelona, Spain \\
                       $and$                        \\
I.F.A.E.}

\def\thefootnote{\fnsymbol{footnote}}
\pagestyle{empty}
%{\hfill \parbox{6cm}{\begin{center} quant-ph/9812xxx\\
%                                    December 1998
%                     \end{center}}}
\vspace{1.5cm}

\begin{center}
\large{Optimal minimal measurements of mixed states}

\vskip .6truein
\centerline {G. Vidal\footnote{e-mail: guifre@ecm.ub.es}, J. I. Latorre, P. Pascual and R. Tarrach}
\end{center}
\vspace{.3cm}
\begin{center}
\ECM
\end{center}
\vspace{1.5cm}

\centerline{\bf Abstract}
\medskip

The optimal
and minimal measuring strategy is obtained for 
 a two-state system prepared in a mixed state with a
probability given by any isotropic a priori distribution. 
We explicitly construct the specific optimal and
minimal  generalized
measurements, which turn out to be independent of the  a priori probability
distribution, obtaining the best guesses
for the unknown state as well as a closed expression for the maximal mean averaged fidelity. 
We do this for up to three copies of the unknown state in a way which leads to the 
generalization to any number of copies, which we then present and prove.

\newpage
\pagestyle{plain}

\section{Introduction}
\renewcommand{\theequation}{1.\arabic{equation}}
\setcounter{equation}{0}

A measurement allows us to extract 
 only a small amount of the information needed to specify a quantum state.
If our preparing device produces several identical copies of the unkown
state, then measurements allow to extract more information, although
only in the limit of infinitely many copies 
we acquire complete knowledge of the unknown quantum state. Performing
an optimal measurement, the one which extracts the maximal possible
amount of information about the state, and among these a minimal
measurement, the one with the minimal number of outcomes, is always a
priority, especially if the process leading to the state is rare or costly. It is also the broad subject of this paper.

 There are two aspects which significantly quantify the difficulty of
the problem. One of them is the dimension of the Hilbert space which
corresponds to the physical system we are considering. We will take the
lowest one, two. The second is the a priori probability distribution
function of the unkown state. If  the state is known to be pure,
 the problem has been solved 
\cite{MP,DE,LPT}. The averaged, mean fidelity of the optimal measurements
performed on $N$ copies of a pure state is \cite{MP}
\begin{equation}
\label{fmax}
\overline {F}^{(N)}_{max} (pure)= {{N+1}\over {N+2}}
\end{equation}
and the minimal measurements correspond, for $N=1$ to $5$, to \cite{LPT}
\begin{equation}
\label{nNmin}
n^{(N)}_{min} (pure)= 2, 4, 6, 10, 12 
\end{equation}
outcomes. The aim of this paper is to solve this problem when we
enlarge
the a priori probability distribution function to include mixed states.
More specifically, when one assumes that it is isotropic and otherwise
arbitrary, but known.

On the other hand, the difficult and  heavily discussed issue 
about which is the abolutely unbiased probability distribution in the
space of density matrices is not settled, and it might even
not have an unbiased solution. In any case an unbiased distribution will
be isotropic in the three-dimensional Poincar\'e sphere covered by the
Bloch vector which parametrizes the unknown density matrix and thus our
results will be valid for any author's prefered candidate for an unbiased
probability distribution. We will not 
discuss  this issue further.
\bigskip

Let us now outline the strategy defining  optimal minimal measurements. 
We consider the simplest possible quantum system, a
two-state system. It might be the spin of an electron, the polarization
of a photon, an atom at very low temperatures so that only the two
lowest hyperfine states matter, a linearly trapped ion for which only
the ground and the first excited vibrational states are important, etc.
This state is described by a $2$ x $2$ density matrix
\begin{equation}
\label{rhob}
\rho (\vec {b})={1\over 2} \left(I + \vec {b}\cdot \vec {\sigma}\right)
={1+b\over 2} \mid \hat {b}\rangle\langle\hat {b} \mid
+ {1-b\over 2}\mid -\hat {b}\rangle\langle-\hat {b} \mid, \null\hskip 1cm
b\equiv \mid \vec {b}\mid \leq 1\ ,
\end{equation}
where $\vec {b}$ is the Bloch vector and $\mid \hat {b}\rangle$ and
$\vert -\hat {b}\rangle$ are the
eigenstates of $\rho (\vec {b})$. These
density matrices  are prepared according to a known, isotropic, a 
priori probability distribution function
given by
\begin{equation}
\label{fb}
f (b) \geq 0 \qquad,\qquad 
4\pi \int ^1_0 db\ b^2\ f(b) = 1\ .
\end{equation}

We will  analyze the generalized measurements performed on the state
corresponding to $N$ copies of $\rho (\vec {b})$, that is,
$\rho(\vec {b})^{\bigotimes N}$, and determine which ones are optimal.
There are two aspects to an optimal measurement: which are the positive
operators correlated to the different outcomes, and which are the
guesses which one makes, given an outcome, about the unkown state (which we shall 
call $\tilde{\rho_i}$).
Optimal measurements have to answer both questions by demanding that the
guesses on average lead to the highest fidelity estimation of $\rho (
\vec {b})$, after averaging over the known probability distribution
function $f(b)$.  We will then determine which of these optimal
measurements are minimal, i.e. have the minimal number of outcomes. For
more than one copy, $N> 1$, measurements may be collective and thus may
involve entanglement. We will have something to say also about the
relation between optimality and entanglement. The role of cloning as
part of an optimal measurement will also be studied. We will also show that for more than two
copies optimal measurements which are minimal are not complete, i.e.
they involve positive operators with rank larger than one (and, yet, are
optimal!).

These are the main issues which will be presented for
$N=1$ to $3$ copies in the next three sections. In  sect. 5 we present and
prove our general results for any $N$.
The last section
briefly recollects our findings and conclusions.

\section{$N=1$}
\renewcommand{\theequation}{2.\arabic{equation}}
\setcounter{equation}{0}

Let us start with one single copy of $\rho$, $N=1$,  and use 
this example to present some of the 
systematics of our approach.
 
 We will first 
perform a generalized measurement [4] on $\rho (\vec {b})$ with $n$
outcomes, given by the operator sum decomposition
\begin{equation}
\label{sum}
\sum^n_{i=1} A_i^\dagger A_i\equiv \sum^n_{i=1} c^2_i
\rho_i= I \qquad 
,\qquad \rho_i=\rho^\dagger_i\geq 0 \qquad , \qquad {\rm Tr} \rho_i= 1 
\end{equation}
which implies
\begin{equation}
\label{sumni}
\sum^n_{i=1} c^2_i= 2\qquad,\qquad
\sum^n_{i=1} c^2_i \vec {s}_i=0 \ ,
\end{equation}
where $\vec {s}_i$ is the Bloch vector of $\rho_i$. If the outcome $i$ is
obtained, which happens with probability
\begin{equation}
\label{prob}
c^2_i \ {\rm Tr} \left(\rho(\vec {b}) \rho_i\right)
= c^2_i{1 \over 2} \left(1+\vec {b}\cdot\vec {s_i}\right)\ ,
\end{equation}
one proposes $\tilde {\rho_i}$ as a guess for the unkown state $\rho(\vec
{b})$. The fidelity, i.e. the measure of the  goodness for 
a proposed guess,
 is quantified by \cite{JOZSA}
\begin{equation}
\label{rhorhoi}
F\left(\rho, \tilde {\rho_i}\right) \equiv \left({\rm Tr}
 \sqrt {\rho^{1/2} \tilde {\rho}_i
\rho^{1/2}}\right)^2 = {1\over 2} \left(1 + \vec {b} \cdot \vec {r}_i
 + \sqrt {1 - b^2}
\sqrt {1-r^2_i}\right)\ , 
\end{equation}
where $\vec {r}_i$ is the Bloch vector of $\tilde {\rho}_i$. Thus, the
fidelity averaged over all outcomes is
\begin{equation}
\label{averagefid}
F^{(N=1)} (\rho) \equiv {1\over 4}\ \sum^n_{i=1} c^2_i \ 
\left(1+\vec {b}\cdot \vec
{s}_i\right) \left(1+
\vec {b}\cdot \vec {r}_i + \sqrt {1- b^2} \sqrt {1-r^2_i}\right)\ ,
\end{equation}
where the superscript reminds us that we are dealing with only one copy.
From here the mean fidelity, i.e. the fidelity averaged over all unkown
states $\rho(\vec {b})$ weighed with the known probability
distribution function $f(b)$, is readily obtained
\begin{eqnarray}
\nonumber
\overline {F}^{(N=1)} &\equiv& \int d \Omega \int ^1_0 db\ b^2\ f (b)
F^{(N=1)}(\rho)\\
\label{fbarone}
&=&
\pi \int ^1_0\ db\ b^2\ f(b) \sum^n_{i=1} c^2_i \left(1
 + {b^2\over 3} \vec
{s}_i\cdot \vec {r}_i + \sqrt {1-b^2} \sqrt {1-r^2_i}\right) \ .
\end{eqnarray}
With the notation
\begin{equation}
\label{notation}
I_\alpha \equiv 4 \pi \int^1_o db \ b^2 f(b) \left({1-b^2 \over 4}
\right)^\alpha\qquad ,\qquad
I_0=1\ ,
\end{equation} 
(note that $I_\alpha-4 I_{\alpha+1}\ge 0$) 
the averaged fidelity reads
\begin{equation}
\label{fbaroneagain}
\overline {F}^{(N=1)}={1\over 4} \sum^n_{i=1}c^2_i \left(1 + {1\over 3} (1 -
4I_1)\ \vec {s}_i\cdot \vec {r}_i + 2 I_{1/2} \sqrt {1-r^2_i}\right) \ .
\end{equation}

We have now to settle which is  the 
best guess for the unknown initial state based on the result of our measurement,
 that is the proposed $\tilde {\rho_i}$ which leads to the
highest mean fidelity. Let us first dispose of the case $4I_1=1$, which corresponds only to $f(b)={1\over 4\pi
b^2}\lim_{\epsilon \rightarrow 0} \delta (b-\epsilon)$, $\epsilon > 0$. It implies a vanishing Bloch vector and thus $\rho(\vec
{b})= {1\over 2}I$, the completely random state. Since the
unknown state is necessarily the completely random state, the state is
known without performing any measurement whatsoever. We will thus always
assume $4 I_1<1$, and only use $4I_1=1$ as a check-up of our results. Then from eq. (\ref{fbaroneagain}) maximization implies that the best guess
corresponds to
\begin{equation}
\label{vecri}
\vec {r}_i = {(1-4I_1)\vec {s}_i\over \sqrt {36 I^2_{1/2} + (1-4I_1)^2
s_i^2}} \ .
\end{equation}

Notice that $\tilde {\rho_i} \not= \rho_i$, but $\tilde {\rho}_i$ is a known
function of $\rho_i$, as its coefficients depend only functionally on
$f(b)$. As $f(b)$ is known, eq. (\ref{vecri}) determines the optimal guess in
terms of $\rho_i$. Substituting one obtains
\begin{equation}
\label{maxri}
{\rm max}_{\vec {r}_i} \overline {F}^{(N=1)}\equiv \overline {F}^{(N=1)}_m=
{1\over 4} \sum ^n_{i=1} c^2_i \left(1 + {1\over 3} \sqrt {36 I^2_{1/2} +
(1-4I_1)^2 s^2_i}\right) º . 
\end{equation}
We now have to determine the best measuring strategy, the one which
leads to the largest possible fidelity. It is obviously given by
$s_i=1$, i.e. by outcomes associated with rank-one projectors, and
gives
\begin{equation}
\label{maxsi}
{\rm max}_{\vec {s}_i} \overline {F}^{(N=1)}_m
= \overline {F}^{(N=1)}_{max}=
{1\over 2} \left( 1 + {1\over 3}
\sqrt {36 I^2_{1/2} + (1-4I_1)^2}\right)\ .
\end{equation}
This is our result for one single copy of the physical system in state
$\rho(\vec {b})$ with a priori probability distribution $f(b)$.

Notice that we have found that  optimal
measurements require necessarily an operator sum decomposition in terms of rank-one
projectors. It is of course obvious that one can always perform an
optimal measurement with rank-one projectors. Suppose, for instance,  that we have some optimal
operator sum decomposition with one operator of rank two, say $\rho_i$. Then
from its spectral decomposition
\begin{equation}
\label{rhoi}
\rho_i=p_i\vert \rho_{i1}\rangle\langle
\rho_{i1}\vert + (1-p_i)\vert \rho_{i2}\rangle\langle\rho_{i2}\vert\ ,
\end{equation}
and from eq.(\ref{prob})
\begin{equation}
\label{cirhoi}
c^2_i {\rm Tr} (\rho(\vec {b})\rho_i)=c^2_ip_i {\rm Tr}\left(\rho(\vec
{b})\vert \rho_{i1}\rangle\langle\rho_{i1}\vert\right)+c^2_i(1-p_i) {\rm Tr} 
\left(\rho(\vec {b})\vert
\rho_{i2}\rangle\langle\rho_{i2}\vert\right)\ ,
\end{equation}
it is clear that taking as the guess for $\rho$ for both outcomes  associated to
$\vert \rho_{i1}\rangle$ and $\vert \rho_{i2}\rangle$ 
precisely $\tilde {\rho}_i$, one can
trade $\rho_i$ for its two rank-one eigenprojectors, having thus a
measurement with only rank-one projectors. This result can be trivially 
generalized to $N$ copies and is of course well-kown [6]. We will use it
without futher comments in obtaining $\overline {F}^{(N)}_{max}$, but it
does not allow to analyze optimal measurements which are minimal, which  will
need a separate treatement.

In the case we are considering here, $N=1$, the outcomes
are necessarily associated to rank-one operators and thus, from eq.
(\ref{sumni}), a minimal optimal measurement requires two outcomes,
$n^{(N=1)}_{min}=2$. This corresponds to a standard von Neumann
measurement, which  is a result unique for $N=1$. For $N>1$ optimal
measurements are generalized measurements.

A limit of interest corresponds to considering 
pure states, which is obtained by taking $f(b)={1\over {4\pi b^2}}
lim_{b_0\rightarrow 1} \delta (b-b_0)$, $b_0<1$.
It follows that $\overline {F}^{(N=1)}_{max} (pure)=
{2\over 3}$, which is the known result given in eq. (\ref{fmax}). Notice that
in this case $\tilde {\rho_i}=\rho_i$ and thus the guess is precisely the pure
state corresponding to the projector, while we have found that for mixed
states the guess $\tilde \rho_i$ is  a mixed state, different, though
related, to the pure state corresponding to the projector. This
is a new feature of optimal measurements.
The two guesses correspond to two points in the interior of the
Poincar\'e sphere and symmetric with respect to its center.
In the other
extreme, discussed after eq. (\ref{fbaroneagain}),
 when one knows that $\rho(\vec
{b})$ is the completely random state, we obtain $\overline
{F}^{(N=1)}_{max} (random)=1$, as it should. One could think that
minimizing $\overline {F}^{(N=1)}_{max}$ with respect to $f(b)$ would lead
to $2\over 3$, as pure states cover the border of the Poincar\'e sphere
and thus maximize the naive distance between the states. This is not so. The
probability distribution function $f(b)={1\over {40 \pi b^2}} (\delta
(b)+9 \delta (b-1))$ gives $\overline {F}^  {(N=1)}_{max}={1\over 2}
(1+{1 \over \sqrt {10}})<{2\over 3}$ and we believe it to be the absolut
minimum.\par

\section{$N=2$}
\renewcommand{\theequation}{3.\arabic{equation}}
\setcounter{equation}{0}

We will now study the situation in which two copies of
the unknown state $\rho(\vec {b})$ are available, i.e. we have the
state $\rho(\vec {b})\otimes \rho (\vec {b})$.
As we shall see, collective measurements appear here for the first time.

 Notice that defining the
exchange operator $V$ by
\begin{equation}
\label{defv}
V \vert\varphi\rangle\otimes\vert\psi\rangle=\vert\psi\rangle
\otimes\vert\varphi\rangle,\qquad
V=V^\dagger=V^{-1}\ ,
\end{equation}
we have the following exchange invariance
\begin{equation}
\label{vrhorhov}
V(\rho \otimes \rho) V=\rho\otimes \rho\ . 
\end{equation}
We will consider generalized measurements for which outcomes correspond
to rank-one projectors, as our purpose now is to build an optimal
measurement. Thus the operator sum decomposition will be written as
\begin{equation}
\label{defi}
\sum^n_{i=1} c^2_i \vert \psi_i\rangle\langle\psi_i\vert= I, \null\hskip 1cm
\vert\psi_i\rangle\in C^2\otimes C^2 \ .
\end{equation}
Given one decomposition one can obtain other decompositions as follows.
First, obviously,
\begin{equation}
\label{defiv}
\sum^n_{i=1} c^2_i\ V \vert\psi_i\rangle\langle\psi_i\vert V=I\ .
\end{equation}
Then, introducing the eigenstates of $V$ built from $\vert\psi_i\rangle$
 and
$V\vert\psi_i\rangle$,
\begin{equation}
\label{psipm}
\vert\psi_i\rangle_\pm\equiv {1\over \sqrt {2} \sqrt 
{1\pm\langle\psi_i\vert
V\vert\psi_i\rangle}} (\vert\psi_i\rangle\pm V\vert\psi_i\rangle)\ ,
\end{equation}
and, as 
\begin{equation}
\label{psipsivv}
\vert\psi_i\rangle\langle\psi_i\vert + 
V\vert\psi_i\rangle\langle\psi_i\vert V=
(1+\langle\psi_i\vert V \vert \psi_i\rangle)\
 \vert \psi_i\rangle_+{}_+\langle
\psi_i\vert
+(1-\langle\psi_i\vert V\vert \psi_i\rangle)\ \vert 
\psi_i\rangle_-{}_-\langle\psi_i\vert \ ,
\end{equation}
we have another decomposition
\begin{equation}
\label{decomp}
{1\over 2} \sum^n_{i=1} c^2_i\left((1+\langle\psi_i\vert V\vert\psi_i\rangle
)\ \vert
\psi_i\rangle_+{}_+\langle  \psi_i\vert + (1-\langle\psi_i\vert V\vert
\psi_i\rangle)\ \vert\psi_i \rangle_-\ {}_-\langle \psi_i \vert\right)=I
\ .
\end{equation}
If the decomposition eq. (\ref{defi}) corresponds to an optimal
measurement, so does eq. (\ref{psipm}) just recalling eq. (\ref{vrhorhov})
 and using the same guesses. Furthermore, as the probability of the {\it
i}-th
outcome is the sum of the probabilities of the $i_+$ and $i_-$ outcomes of
the decomposition of eq. (\ref{decomp}),
\begin{eqnarray}
\nonumber
\label{cirhodos}
c^2_i\ \langle\psi_i\vert \rho\otimes \rho \vert \psi_i\rangle
&=&{c^2_i\over 2}
(1+\langle\psi_i\vert V\vert\psi_i\rangle) \ {}_{+}\langle
\psi_i\vert \rho\otimes
\rho\vert\psi_i\rplus\\
&+&{c^2_i\over 2} (1-<\psi_i\vert V\vert \psi_i>){}_-\langle\psi_i\vert
\rho\otimes \rho\vert\psi_i\rangle_- 
\end{eqnarray}
it is enough to associate again the same guess to the $i_+$ and $i_-$
outcomes to make the measurement of eq. (\ref{decomp}) optimal too. Thus
optimal measurements can always be obtained by projecting on eigenstates
of $V$.

An equivalent way of presenting these results, and
which will be more convenient for $N>2$, is based on the identity
\begin{equation}
\label{vssi}
V= \vec {S}^2 - I 
\end{equation}
relating the exchange operator with the square of the total spin operator,
\begin{equation}
\label{ssigmasigma}
\vec {S}\equiv{1\over 2} (\vec {\sigma} \otimes I + I \otimes \vec {\sigma}).
\end{equation}
Eq. (\ref{vrhorhov}) now reads
\begin{equation}
\label{comss}
\left[\vec {S}^2, \rho \otimes \rho\right]=0 
\end{equation}
and our previous results allow to write eq. (\ref{defi}) as
\begin{equation}
\label{resid}
\vert \sigma\rangle\langle\sigma\vert + \sum^{n-1}_{i=1} c^2_i \ \vert
\tau_i \rangle\langle\tau_i\vert=I 
\end{equation}
where $\vert \sigma\rangle$
 is the singlet or antisymmetric state, and $\vert
\tau_i\rangle$ are triplet or symmetric states. This is an important result. It states that decomposing
the Hilbert space of the two copies A and B into a direct sum of
eigenspaces of $\vec {S}^2$,
\begin{equation}
\label{hilbert}
{\cal H}^{(N=2)}\equiv {\cal H}_A \otimes
 {\cal H}_B= E_0 \oplus E_1\ ,
\end{equation}
where $E_s$ corresponds to the eigenvalue $s(s+1)$ of $\vec {S}^2$, it
is enough to find optimal measurements in each of the spin eigenspaces
for obtaining an optimal measurement in the whole space. The
generalization of this result to $N>2$ will be essential. It will then
also be convenient to use both spin and exchange invariances
simultaneously.

We are ready to resume our general strategy for performing optimal measurements. 
First, the probability that the
outcome corresponds to the singlet state is
\begin{equation}
\label{probsigsig}
\langle\sigma\vert \rho\otimes \rho\vert\sigma\rangle
={{1-b^2}\over 4}\ .
\end{equation}
For the triplet states we have found it convenient to
use the Hilbert-Schmidt parametrization
\begin{eqnarray}
\label{triplet}
\nonumber
\vert \tau_i \rangle\langle\tau_i\vert&=&{1\over 4}\left(I\otimes I+\vec
{t}_i\cdot\vec {\sigma}\otimes I+ I\otimes \vec {t}_i\cdot\vec
{\sigma}+\hat {t}_i\cdot \vec {\sigma}\otimes \hat {t}_i\cdot\vec {\sigma}
\right.
\\
&+&\left.\sqrt {1-t^2_i}\ (\hat {u}_i\cdot \vec {\sigma}\otimes \hat {u}_i\cdot
\vec {\sigma} - \hat {v}_i\cdot \vec {\sigma}\otimes \hat {v}_i\cdot\vec
{\sigma})\right)
\end{eqnarray}
where $\hat {t}_i$, $\hat {u}_i$ and $\hat {v}_i$
are $n-1$ triads of orthonormalized
vectors. Notice that $\vec {t}_i$ is the Bloch vector of the reduced
density matrix
\begin{equation}
\label{subtraces}
{\rm Tr}_A\vert \tau_i\rangle\langle\tau_i\vert= 
{\rm Tr}_B\vert\tau_i\rangle\langle\tau_i\vert={1\over
2}\left(I+\vec {t}_i\cdot \vec {\sigma}\right)\equiv \rho_i\ ,
\end{equation}
where we use subscripts $A$ and $B$ to earmark the Hilbert space over which the trace is performed. Furthermore from eq. (\ref{resid})
 we have
\begin{equation}
\label{restric}
\sum^{n-1}_{i=1} c^2_i= 3\qquad,\qquad
\sum^{n-1}_{i=1} c^2_i \vec {t}_i=0 \ 
\end{equation}
and further restrictions on $\hat u_i$, $\hat v_i$ and $\hat t_i$ which 
will not be needed here.
The probability that the outcome corresponds to $\vert \tau_i\rangle$ is
\begin{equation}
\label{probtaui}
c^2_i\ \langle\tau_i\vert \rho\otimes \rho\vert\tau_i\rangle
={c^2_i\over 4}
\left(1+2\ \vec
{b}\cdot \vec {t}_i+(\vec {b}\cdot\hat {t}_i)^2+\sqrt
{1-t^2_i}\ ((\vec {b}\cdot \hat {u}_i)^2-(\vec {b}\cdot \hat
{v}_i)^2)\right)\ .
\end{equation}
Once outcome $i$ is obtained one proposes $\tilde {\rho}_i$ as a guess of
the unknown state $\rho(\vec {b})$. From eq. (\ref{rhorhoi})
 one obtains for the
fidelity averaged over outcomes
\begin{eqnarray}
\label{fdosrho}
\nonumber
F^{(N=2)}(\rho)&=&{1\over 8} (1-b^2)\left(1+\vec {b}\cdot\vec
{r}_n+\sqrt {1-b^2} \sqrt {1-r_n^2}\right)
\\
\nonumber
&+&{1\over 8} \sum^{n-1}_{i=1} c^2_i\left(1+2\vec {b}\cdot\vec
{t}_i+(\vec {b}\cdot\hat {t}_i)^2+\sqrt {1-t^2_i}((\vec
{b}\cdot \hat {u}_i)^2-(\vec {b}\cdot\hat {v}_i)^2)\right)
\\
&&\left(1+\vec {b}\cdot\vec {r}_i+\sqrt {1-b^2} \sqrt
{1-r^2_i}\right)\ .
\end{eqnarray}
The mean fidelity is obtained after averaging over the state space with
the probability distribution function and reads
\begin{eqnarray}
\label{fdosmax}
\nonumber
\overline {F}^{(N=2)}&=&{1\over 2}\left(I_1+2I_{3/2}\sqrt {1-r^2_n}\right)
\\
&+&{1\over 6} \sum^{n-1}_{i=1}c^2_i \left(1-I_1+{1\over 2}(1-4I_1)\vec
{t}_i\cdot\vec {r}_i+2 (I_{1/2}- I_{3/2})\sqrt {1-r^2_i}\right)\ .
\end{eqnarray}
From here the best guesses are readily obtained
\par
\begin{center}
$r_n=0$ (except for $f(b)={1\over 4\pi}\delta (b-1)$
when $r_n$ is not
determined)
\end{center}
\begin{equation}
\label{rsubi}
\vec {r}_i = {(1-4I_1)\over \sqrt {16(I_{1/2}-I_{3/2})^2+(1-4I_1)^2
t^2_i}}\vec {t}_i\qquad i=1,\dots,n-1 
\end{equation}
As before for $N=1$, again $\tilde {\rho}_i \not=\rho_i$ is a function of $\rho_i$, in
fact a mixture of $\rho_i$ and the completely random state. Substituting
the best guesses we obtain
\begin{equation}
\label{fdosm}
\overline {F}^{(N=2)}_m={1\over 2} I_1+I_{3/2}+{1\over 6}\sum^{n-1}_{i=1}
c^2_i
\ \left(1-I_1+{1\over 2}\sqrt {16 (I_{1/2}-I_{3/2})^2+(1-4I_1)^2
t^2_i}\right) \ .
\end{equation}
The best measurement strategy is obtained for $t_i=1$, so that $\rho_i$ is
a pure state and $\vert \tau_i\rangle$ is a product state, without
entanglement. This is a reasonable result, since $\rho\otimes \rho$ has
neither entanglement nor classical correlations, so that it would be
surprising that projecting on entangled states would lead to an optimal
measuring strategy. Notice also that this result of no entanglement,
which we will reencounter later for $N>2$, is independent of $f(b)$. In fact,
once the specification of the operator sum decomposition does not depend
on $f(b)$, it has to correspond to an optimal measurement strategy valid
for pure states. But this is known [1, 2] to precisely require product
states. For the
singlet, which is a maximally entangled state, there are no
alternatives, and thus the previous argument is irrelevant. The final
result is 
\begin{equation}
\label{fmaxdos}
\overline {F}^{(N=2)}_{max}={1\over 2}+ I_{3/2}+{1\over 4}\sqrt {16
(I_{1/2}-I_{3/2})^2+(1-4I_1)^2}
\end{equation}

This final result reproduces the known limits. Indeed, 
the pure state result of eq. (\ref{fmax}) is readily obtained from eq.
(\ref{fmaxdos}), when $f(b)={1\over 4\pi} \delta (b-1)$. Also for the
completely random state $\overline {F}_{max}^{(2)} (random)=1$. One can
also check from the comparison of $(\overline {F}^{(i)}_{max}-
{1\over 2})^2$ for $i=1$ and $2$ that, as it should,
\begin{equation}
\label{fdosfone}
\overline {F}^{(N=2)}_{max} \geq \overline {F}^{(N=1)}_{max}\ .
\end{equation}

Let us now analyze optimal measurements which are
minimal. With the constraints we have been using for obtaining optimal
measurements, i.e. an operator sum decomposition in terms of rank-one
symmetric or antisymmetric projectors, the minimal $n$ is five. This is
because in the 3-dimensional symmetric (triplet) space a resolution of
the identity in terms of symmetric product states needs four of them [3],
which together with the singlet makes five. When the unkown state is
known to be pure, the outcome corresponding to the singlet never
happens, and one can do with just four projectors. Let us now prove that
one cannot do with less.

Suppose we have an optimal measurement such that one of
the rank-one projectors of its operator sum decomposition, $\vert
\psi\rangle\langle\psi\vert$, with associated best guess $\tilde {\rho}$,
is not symmetric nor antisymmetric. Obviously the best guess associated to
$V\vert\psi\rangle\langle\psi\vert V$ is also $\tilde {\rho}$. One can then build,
following the arguments of eqs. (\ref{psipm}-\ref{cirhodos})
 an optimal measurement with
$\vert\psi\rangle_+\ {}_+\langle\psi\vert$
 and $\vert\psi\rangle_-\ {}_-\langle\psi\vert$ with
associated best guesses $\tilde {\rho}$ for both of them. But this is
impossible, as we saw that the best guess associated to the
antisymmetric state is the completely random state, while the one
associated to the symmetric state has a non-vanishing Bloch vector (see
eq. (\ref{rsubi}), and thus the best guesses cannot be equal.\par

The very same reasoning forbids to have an optimal
measurement with an operator sum decomposition for which one of the
operators has rank larger than one, as the associated rank-one
projectors which appear in its spectral decomposition will have
necessarily different best guesses. The upshot of all this is that for
$N=2$ minimal optimal measurements correspond to operator sum
decompositions of rank-one symmetric or antisymmetric projectors, and
thus have five outcomes, $n^{(N=2)}_{min}=5$. We will see that for $N>2$
the result that minimal measurements correspond to
rank-one projectors does not hold. Notice that the five guesses
are situated one at the center of the Poincar\'e sphere and the other
four on a concentric shell in its interior forming a regular
tetrahedron.

A related question to which we turn briefly is whether
circumstances exist for which von Neumann measurements can be minimal
and optimal. As $C^2\otimes C^2$ is of dimension four a von Neumann
measurement has four outcomes. We have seen that optimal measurements
with four outcomes only exist when we know that the unknown
state is pure. The question then is if the four triplet states, which
are certainly not orthogonal, can be made orthogonal by adding them
coherently to the singlet state. Notice that these states would not have
a well-defined symmetry, but our previous proof that such states cannot
be part of an optimal measurement fails precisely only for pure states,
as then (cf. first of eq. (\ref{rsubi}))  $r_n$ is arbitrary. It is thus
a legitimate question. Its answer is yes, for $N=2$ \cite{MP}.
 The answer for
$N>2$ is not known.

Let us briefly go back to the situation in which we had
one copy (section 2), and let us clone it with a state-independent
universal quantum cloner [7-11]. The conditions of strong [12]
symmetry and isotropy
of a universal 1-to-2 quantum cloner imply
\begin{equation}
\label{cloner}
\rho(\vec {b}) \rightarrow \rho_c^{(2)} \equiv {1\over 4} (I \otimes I+\eta
(\vec {b}\cdot \vec {\sigma} \otimes I + I \otimes \vec {b} \cdot \vec
{\sigma})+t_{ij} \sigma_i \otimes \sigma_j),\quad t_{ij}=t_{ji}\ ,
\end{equation}
where $\eta$ is the shrinking factor and where $t_{ij}$ depends only on
the vector $\vec {b}$ and the invariant tensor $\delta_{ij}$. Linearity,
which originates in state-independence, and the absence of measurements
in optimal cloning [13] forbids the quadratic dependence
on $b_i$, so that eventually $t_{ij} = t\ \delta_{ij}$. It is also
linearity which allows to clone straightforwardly for $N=1$ a mixed
state by just mixing statistically the clones of the pure states which
realize the mixed state. The values of the real parameters $\eta$ and
$t$ have to be such that $\rho_c^{(2)}$ is a density matrix, i.e. such that
its eigenvalues
\begin{equation}
\label{soletat}
{1\over 4} (1\pm 2 b \eta + t)\quad,\quad {1\over 4} (1 + t)\quad,\quad
 {1\over 4} (1-3t)
\end{equation}
lie between $0$ and $1$. Of course measuring on $\rho_c^{(2)}$ will allow
to learn the most about $\vec {b}$ for the largest $\eta$ possible. This
is precisely what optimal cloning does: $\eta={2\over 3}$ and thus
$t={1\over 3}$. We can now perform an optimal measurement on the
optimal clone $\rho_c^{(2)}$, following closely the study of the $N=2$
case, as $V\rho_c^{(2)} V = \rho_c^{(2)}$. From the following results,
\begin{equation}
\label{rhocdos}
\langle\sigma\vert \rho_c^{(2)} \vert \sigma \rangle=0\qquad,\qquad
\langle\tau_i\vert \rho_c^{(2)} \vert \tau_i\rangle
={1\over 3} \left(1+\vec {b}\cdot
\vec {t}_i\right)\ ,
\end{equation}
the expression equivalent to eq. (\ref{fdosrho}), after dropping an
irrelevant part, is
\begin{equation}
\label{fcdos}
F^{(2)}_c (\rho)= {1\over 6} \sum^{n-1}_{i=1} c^2_i \left(1+\vec {b}\cdot
\vec {t}_i\right) \left(1+\vec {b}\cdot\vec {r}_i + \sqrt {1-b^2} \sqrt {1-r^2_i}\right)
\end{equation}
This expression, together with eq. (\ref{restric}), is identical to eq.
 (\ref{averagefid}), when eq. (\ref{sumni}) is recalled. We thus recover the result
of eq. (\ref{maxsi}). In words, optimal cloning can be part of an optimal
measurement. As a byproduct we have checked that indeed $\rho_c^{(2)}$
with $t={1\over 3}$ and $\eta={2\over 3}$ is the optimal clone of
$\rho (\vec {b})$.

Notice also the result shown in the first of eq.
(\ref{rhocdos}): the optimally cloned state lives in the triplet space.
This is not surprising, as the singlet space cannot carry any
information about the original cloned state.

\section{$N=3$}
\renewcommand{\theequation}{4.\arabic{equation}}
\setcounter{equation}{0}

Consider now three copies of the unknown state,
$\rho\otimes \rho\otimes \rho$. Let us recall its exchange invariances
\begin{equation}
\label{rhorhorho}
\left[V_{AC}, \rho\otimes \rho\otimes \rho\right]=
\left[V_{BC}, \rho\otimes \rho\otimes \rho\right]=0,
\end{equation}
where $A, B, C$ are the subindices
labeling the copies which are exchanged, and its spin
invariances
\begin{equation}
\label{spinrhorhorho}
\left[\vec {S}^2, \rho\otimes \rho\otimes \rho\right]=
\left[\vec {S}^2_{AB},
\rho\otimes \rho\otimes \rho\right]=0,
\end{equation}
where the partial and total spin operators are
\begin{equation}
\label{sab}
\vec {S}_{AB} \equiv {1\over 2} \left(\vec {\sigma}\otimes I\otimes I + I\otimes
\vec {\sigma} \otimes I\right) 
\qquad,\qquad
\vec {S} \equiv \vec {S}_{AB} + {1\over 2}I\otimes I \otimes \vec {\sigma}.
\end{equation}
The first of eq. (\ref{spinrhorhorho}) is obvious if one convinces
oneself first that
\begin{equation}
\label{pthree}
\rho\otimes\rho \otimes \rho = p_3 \left(\vec {S}\cdot \vec {b}\
\right),
\end{equation}
where $p_N (x)$ is a polynomial in $x$ of degree $N$. The second of eq.
(\ref{spinrhorhorho}) follows then immediately.
With the adequate generalizations in going from $N=2$ to $N=3$, it can be seen that in order to obtain optimal
measurements it is enough to consider operator sum decompositions whose
elements are of rank one and project on states which are simultaneous
eigenstates of $\vec {S}^2$ and $\vec {S}^2_{AB}$. Moreover these states should again be eigenstates of $\vec {S}\cdot \hat {n}$ for some $\hat {n}$ with maximal
eigenvalue. Using the
notation\  $\vert s, s_{AB}, \hat {n}\rangle$, this
leads immediately to the following states in terms of which the optimal
operator sum decomposition can be built:
\begin{eqnarray}
\label{decompositio}
\nonumber
&&\vert {3\over 2}, 1, \hat {n}\rangle=\vert \hat {n}\rangle\vert \hat {n}\rangle\vert \hat {n}\rangle \\
\nonumber
&&\vert {1\over 2}, 0, \hat {n}\rangle= \vert \sigma \rangle \vert \hat {n}\rangle\\
&&\vert {1\over 2}, 1, \hat {n}\rangle={1\over \sqrt {3}} (V_{AC}- V_{BC})
\vert \sigma\rangle \vert \hat {n}\rangle . 
\end{eqnarray}
The first state also corresponds to the completely symmetric representation
of the permutation group generated by the exchange operators, and the
other two correspond to the two-dimensional mixed symmetry
representation of the same group. We may recall from ref. \cite{LPT} that six
states of the type of the first one of eq. (\ref{decompositio}) pointing into the
six directions of the vertices of a regular octahedron resolve the
identity in the four-dimensional maximal spin space, $
s={3\over 2}$.  Therefore, we obtain the following optimal operator sum
decomposition
\begin{eqnarray}
\label{sumdecomp}
\nonumber
{2\over 3}\sum^6_{i=1}
\left(\vert \hat {n}_i\rangle \langle\hat {n}_i\vert\right)^{\otimes 3}+
\vert \sigma\rangle \langle\sigma\vert \otimes
\vert \hat {n}\rangle \langle\hat {n}\vert + \vert \sigma
\rangle \langle\sigma\vert\otimes
\vert -\hat
{n}\rangle \langle-\hat {n}\vert&&\\
\nonumber
+{1\over 3} (V_{AC} - V_{BC}) \vert\sigma\rangle \langle
\sigma\vert\otimes\vert \hat {n}\rangle \langle\hat {n}\vert
(V_{AC}-V_{BC})&&\\
+{1\over 3} (V_{AC} - V_{BC}) \vert\sigma\rangle \langle
\sigma\vert\otimes \vert -\hat {n}\rangle \langle-\hat {n}\vert
(V_{AC}-V_{BC})&=&I\ .
\end{eqnarray}
This result recalls the decomposition into eigenspaces 
$E_{s,s_{AB}}$of $\vec {S}^2$
and $\vec {S}^2_{AB}$,
\begin{equation}
\label{eigenspaces}
{\cal H}^{(N=3)}\equiv {\cal H}_A \otimes {\cal H}_B \otimes {\cal H}_C
=E_{{3\over 2}, 1} \oplus E_{{1\over 2}, 0} \oplus
E_{{1\over 2}, 1}
\end{equation}
and that under permutations $E_{{1\over 2}, 0}$ can be transformed into
$E_{{1\over 2}, 1}$. (Let us note here that the correctness of
eq. (\ref{sumdecomp}) has been confirmed by a brute force
assumption-free computation which we performed in early stages of this work). Because of the isotropy of the 
probability distribution $f(b)$
we just need to compute the following probabilities
\begin{eqnarray}
\label{probabi}
\nonumber
&&\langle\hat {n}\vert\langle\hat {n}\vert \langle\hat {n}\vert \rho\otimes \rho\otimes \rho\vert
\hat {n}\rangle\vert \hat {n}\rangle \vert \hat {n}\rangle = 
\langle\hat {n}\vert \rho \vert \hat
{n}\rangle^3= {1\over 8} \left(1+\vec {b}\cdot \hat {n}\right)^3\\
\nonumber
&&\langle\sigma \vert \langle\hat {n}\vert \rho\otimes 
\rho\otimes \rho\vert \sigma\rangle\vert \hat {n}\rangle=
\langle\sigma\vert \rho\otimes \rho \vert \sigma\rangle \langle
\hat {n}\vert \rho \vert \hat {n}\rangle
= {1-b^2\over 8} \left(1+\vec {b}\cdot
\hat {n}\right)\\
&&{1\over 3} \langle\sigma \vert \langle\hat {n}\vert (V_{AC}-V_{BC})
 \rho\otimes \rho\otimes \rho
(V_{AC}-V_{BC}) \vert \sigma\rangle\vert \hat {n}\rangle = \langle
\sigma \vert \langle\hat {n}\vert
\rho\otimes \rho\otimes \rho \vert \sigma\rangle\vert \hat {n}\rangle
\end{eqnarray}
where the last expression is obtained  from
\begin{equation}
\label{sigman}
{1\over 3} (V_{AC}-V_{BC})^2 \vert \sigma\rangle \vert \hat {n}\rangle
= \vert \sigma\rangle \vert \hat {n}\rangle \ . 
\end{equation}
Putting all the pieces  together we obtain  (from eq. (\ref{prob}))
\begin{eqnarray}
\label{fthreerho}
\nonumber
F^{(N=3)}(\rho)&=&{1\over 4} (1-b^2) (1+\vec {b}\cdot \hat {n}) \left(1+ \vec
{b}\cdot \vec {r}_m + \sqrt {1-b^2} \sqrt {1-r^2_m}\right)\\
&+&{1\over 4} (1+\vec {b} \cdot \hat {n})^3 \left(1+\vec {b}\cdot \vec {r}_s+
\sqrt {1-b^2} \sqrt {1-r^2_s}\right) \ ,
\end{eqnarray}
where $\vec {r}_m$ and $\vec {r}_s$ are the Bloch vectors of the
proposed guesses of $\rho$ corresponding to the mixed symmetry and
completely symmetric projectors respectively. Angular integration over
$\hat {b}$ leads to
\begin{eqnarray}
\label{fbarthree}
\nonumber
\overline {F}^{(N=3)}= {1\over 2} &+& {1\over 3} (I_1 - 4I_2) \hat {n}
\cdot \vec {r}_m+ 2I_{3/2} \sqrt {1-r^2_m}\\
&+&(I_{1/2} - 2I_{3/2}) \sqrt {1-r^2_s} + {1\over 10} (3-14 I_1+8I_2) \ \hat
{n}\cdot \vec {r}_s\ , 
\end{eqnarray}
from which the optimal guesses are obtained for
\begin{eqnarray}
\label{rmrs}
\nonumber
&\vec {r}_m& = {(I_1 - 4I_2)\over \sqrt {36 I^2_{3/2} + (I_1-4 I_2)^2}}
\hat {n}\\
&\vec {r}_s&= {3-14 I_1 + 8 I_2 \over \sqrt {100 (I_{1/2} - 2 I_{3/2})^2
+ (3 -14 I_1 + 8 I_2)^2}} \hat n\ .
\end{eqnarray}
Substitution into eq. (\ref{fthreerho})
 leads to our final result for $N=3$,
\begin{eqnarray}
\label{fthreemax}
\nonumber
\overline {F}^{(N=3)}_{max}={1\over 2} &+& {1\over 3} \sqrt {36 I^2_{3/2}
+ (I_1 - 4I_2)^2}\\
 &+& {1\over 10} \sqrt {100 (I_{1/2} - 2I_{3/2})^2 +
(3-14I_1 + 8I_2)^2}\ .
\end{eqnarray}
This result reproduces the pure state result of eq. (\ref{fmax}) and gives 1 for the
completely random state, as in previous cases.

Let us finally come to those optimal measurements which
are minimal. Up to now we have an optimal measurement with 10 outcomes.
Remember that the only possibility of grouping together two rank-one
projectors of the operator sum decomposition happens when the two
different outcomes correspond to the same guess. Now from our results it
is clear that this happens twice, that is the guesses corresponding to
the 7-{\it th} and 9-{\it th} terms of eq. (\ref{sumdecomp}) are the same and given by the
first of eq. (\ref{rmrs}), and the ones corresponding to the 8-{\it th}
 and 10-{\it th}
terms of eq. 
(\ref{sumdecomp}) are also the same and given by the first of eq.
(\ref{rmrs}), but with opposite sign. Thus the minimal optimal
measurement has eight outcomes, $n^{(3)}_{min} = 8$.
The corresponding positive operators
${\cal O}_{N,s,i}$ and guesses
$\rho_{N,s,i}$ for $N=3$ are
({\sl cf.} eq. (\ref{sumdecomp})) six for the space $E_{3/2,1}$:
\begin{equation}
\label{othree}
{\cal O}_{3,3/2,i}={2\over 3} \vert \hat n_i\rangle\langle \hat n_i\vert
{}^{\otimes 3}\qquad,\qquad \rho_{3,3/2,i}={1\over 2}
\left( I+ r_s\ \hat n_i\cdot\vec\sigma\right)\ , 
\end{equation}
and two for the space $E_{1/2,0}\oplus E_{1/2,1}$:
\begin{eqnarray}
\label{oonehalf}
\nonumber
{\cal O}_{3,1/2,1}&=&\vert\sigma\rangle\langle \sigma\vert\otimes
\vert \hat n\rangle\langle\hat n \vert+{1\over 3}
\left(V_{AC}-V_{BC}\right) \vert\sigma\rangle\langle \sigma\vert\otimes
\vert \hat n\rangle\langle\hat n \vert\left(V_{AC}-V_{BC}\right)\\
\nonumber
\rho_{3,1/2,1}&=&{1\over 2}\left(I+ r_m \hat n\cdot\vec \sigma\right)\\
\nonumber
{\cal O}_{3,1/2,2}&=&\vert\sigma\rangle\langle \sigma\vert\otimes
\vert -\hat n\rangle\langle-\hat n \vert+{1\over 3}
\left(V_{AC}-V_{BC}\right) \vert\sigma\rangle\langle \sigma\vert\otimes
\vert -\hat n\rangle\langle-\hat n \vert\left(V_{AC}-V_{BC}\right)\\
\rho_{3,1/2,2}&=&{1\over 2}\left(I- r_m \hat n\cdot\vec \sigma\right)\ .
\end{eqnarray}

 This is the first
time in which a minimal optimal measurement has operators of rank two in
its decomposition. The Bloch vectors of the corresponding guesses are
situated on two concentric shells in the interior of the Poincar\'e
sphere.\par

Notice that again the measuring strategy, i.e. eq.
(\ref{sumdecomp}), is independent of $f(b)$ and thus determined actually by what
is known from [1, 2, 3]: for each $s$ the pure state strategy for $2s$
copies is the optimal strategy. This will allow us to prove the general
expression for $\overline {F}^{(N)}_{max}$ and $n^{(N)}_{min}$ for any
$N$ with relative ease in the next section.

\section{ General results for $N > 3$}
\renewcommand{\theequation}{5.\arabic{equation}}
\setcounter{equation}{0}

We will analyze in this section optimal and minimal
generalized measurements when a generic number $N$ of copies of the unknown state are
available. We present here the maximal fidelity $\overline {F}^{(N)}_{max}$
one can obtain on average by performing such collective measurements over
$\rho^{\otimes N}$, together with the minimal 
number $n^{(N)}_{min}$ of outcomes an optimal generalized
measurement
can have. For any $N$ we provide also a generalized measurement
which is both optimal and
minimal. Explicit results for the case $N=4$ are worked out in order to
illustrate the general expressions.\par

We first display our final, general results:
\begin{equation}
\label{fmaxgen}
\overline {F}^{(N)}_{max}= {1\over 2} + \sum^{N/2}_{s=s_0}
{(2s+1)^2\over {N\over 2} +s+1} {N\choose {N\over 2}+s} \sqrt
{g_1(N, s)^2+g_2(N, s)^2}, 
\end{equation}
where
\begin{eqnarray}
\label{ges}
\nonumber
&&g_1 (N, s)\equiv \int d\Omega \int^1_0 \ db\ b^2 f(b) \left({1-b^2\over
4}\right)^{{N+1\over 2}-s} \left({1+b_z\over 2}\right)^{2s},\\
&&g_2(N, s)\equiv \int d\Omega \int^1_0 \ db\ b^2 f(b) \left({1-b^2\over
4}\right)^{{N\over 2}- s} \left({1+b_z\over 2}\right)^{2s} {b_z\over 2}, 
\end{eqnarray}
$b_z$ is the third component of $\vec {b}$
and $s_0$ is $0$ $(1/2)$ for even (odd) $N$. As for $n^{(N)}_{min}$ we
have found that
\begin{equation}
\label{nmingen}
n^{(N)}_{min} = \sum^{N/2}_{s=s_o} n^{(2s)}_{ps}\ ,
\end{equation}
where we define $n^{(N)}_{ps} \equiv n^{(N)}_{min} (pure)$,
$n^{(0)}_{ps}\equiv 1$. For $N=1$ to $5$ this reads (using \cite{LPT})
\begin{equation}
\label{nminpar}
n^{(N)}_{min}=2, 5, 8, 15, 20. 
\end{equation}

For $N>5$ the minimal $n^{(N)}_{ps}$ relies on a conjecture proposed
in [3], and this is therefore also the case of $n^{(N)}_{min}$ for 
$N>5$.\par

For some very specific a priori probability distributions
$f(b)$ this number can be reduced. This, though,  corresponds only to
cases in which there is an accidental degeneracy in the proposed
guesses, as in the case $f(b)={1\over 4\pi}\delta (b-1)$ (pure states).\par

The optimal and minimal generalized measurements
consists of
the following decomposition of the identity operator in the space
${\cal H}^{(N)}=C^{2\otimes N}$ of the $N$ copies in terms of positive
operators ${\cal O}_{N, s, i}$ and the corresponding guesses $\rho_{N,
s, i}$: for each $s \epsilon [s_o, s_o+1, ..., {N\over 2} - 1,
{N\over 2}]$, our optimal and minimal generalized measurement
contains $n^{(2s)}_{ps}$
positive operators of the form
\begin{equation}
\label{ogen}
{\cal O}_{N, s, i}=c^2_{s,i}{(2s+1)\over {N\over 2} + s + 1} {N\choose
{N\over 2}+s} {1\over N!} \sum_{V\epsilon S_N} V\left(\vert \sigma \rangle\langle \sigma
\vert^{\otimes {N\over 2}-s} \otimes \vert \hat {n}_{s,i} \rangle\langle \hat
{n}_{s,i}\vert^{\otimes 2s}\right)V^\dagger\ ,
\end{equation}
where $S_N$ is the group of the $N!$ possible permutations of $N$
elements acting on the Hilbert space of the $N$ copies, and $c^2_{s,i}$
is such that
\begin{equation}
\label{idengen}
\sum^{N/2}_{s=s_0} \sum^{n^{(2s)}_{ps}}_{i=1}{\cal O}_{N, s, i}=I \ .
\end{equation}
The corresponding guesses are
\begin{equation}
\label{guessgen}
\rho_{N, s, i}= {1\over 2}(I + r_{N, s}\ \hat {n}_{s, i}\cdot \vec {\sigma}),
\end{equation}
where
\begin{equation}
\label{rngen}
r_{N, s} = {g_2(N, s)\over \sqrt {g_1 (N, s)^2 + g_2 (N, s)^2}}.
\end{equation}

The $n^{(2s)}_{ps}$ vectors $\hat {n}_{s,i}$ are
distributed according to their counterparts of the $N=2s$ case of
optimal estimation of pure states as described in \cite{LPT}, and the
coefficients $c^2_{s,i}$ satisfy 
\begin{equation}
\label{condgen}
\sum_{i=1}^{n^{(2s)}_{ps}} c^2_{s, i} \ \hat {n}_{s, i}= 0\qquad,\qquad
\sum_{i=1}^{n^{(2s)}_{ps}} c^2_{s, i}=2s+1 \ .
\end{equation}
For $s={1\over 2}, 1, {3\over 2}, {5\over 2}$
 they are independent of $i:c^2_{s, i}
={2s+1\over n_{ps}^{(2s)}}$. All these results are essentially unique.

\bigskip
For $N=4$ our results can be explicitly written as
\begin{eqnarray}
\label{fmaxfour}
\nonumber
\overline {F}^{(N=4)}_{max} = {1\over 2}&+& 2 I_{5/2} + {1\over 6} \sqrt
{\left( 2-11I_1+12I_2\right)^2 + 36 \left(I_{1/2}-3 I_{3/2}+ I_{5/2}\right)^2}\\
&+&{3\over 4} \sqrt {\left(I_1-4I_2\right)^2 + 16 \left(I_{3/2} - I_{5/2}\right)^2} 
\end{eqnarray}
and 
\begin{equation}
\label{nminfour}
n_{min}^{(N=4)}=15.
\end{equation}
The positive operator sum decomposition reads
\begin{equation}
\label{idecompfour}
I={\cal O}_{4,0} + \sum^4_{i=1} {\cal O}_{4, 1, i} + \sum^{10}_{i=1} 
{\cal O}_{4, 2, i}\ ,
\end{equation}
where to the rank-two projector
\begin{equation}
\label{ranktwoop}
O_{4, 0}= {1\over {12}} \sum_{V \epsilon S_4} V\vert \sigma \rangle\langle \sigma
\vert\otimes\vert \sigma \rangle\langle \sigma \vert V^\dagger
\end{equation}
there corresponds the guess
\begin{equation}
\label{guessfour}
\rho_{4,0}={1\over 2}I  \qquad \qquad  (r_{4,0}=0) \ .
\end{equation}
The $4$ rank-three positive operators
\begin{equation}
\label{rankthree}
{\cal O}_{4, 1, i}={3\over {32}} \sum_{V\epsilon S_4} V\vert\sigma
\rangle\langle\sigma\vert\otimes\vert \hat {n}_{1,i}\rangle\langle\hat {n}_{1,i}\vert^{\otimes
2}V^\dagger  \qquad,\qquad i=1,\dots,4
\end{equation}
have associated guesses
\begin{equation}
\label{fourguess}
\rho_{4,1,i}={1\over 2} (I+r_{4,1}\ \hat {n}_{1,i} \cdot \vec
{\sigma}),\hskip 1cm r_{4,1}={I_1 - 4I_2\over \sqrt {(I_1 - 4I_2)^2 +
16 (I_{3/2} - I_{5/2})^2}} 
\end{equation}
(here the $\hat {n}_{1,i}$ are distributed according to a regular
tetrahedron \cite{LPT}), and the $10$ rank-one positive operators
\begin{equation}
\label{tenrankone}
{\cal O}_{4,2,i}=c^2_{s,i}\vert \hat {n}_{2,i} \rangle\langle \hat {n}_{2, i}
\vert^{\otimes 4} \qquad,\qquad 
 i=1,\dots,10
\end{equation}
have associated guesses
\begin{equation}
\label{fourguesses}
\rho_{4, 2, i}={1\over 2} (I+r_{4, 2}\  \hat {n}_{2, i} \cdot \vec
{\sigma}),\qquad r_{4, 2}= {(2-11 I_1 + 12 I_2)^2 \over \sqrt {(2
- 11 I_1 + 12 I_2)^2 + 36 (I_{1/2} - 3I_{3/2} + I_{5/2})^2}}
\end{equation}
(a concrete solution for $\hat {n}_{2,i}$ and $c^2_{2,i}$ is given in
\cite{LPT}).

\bigskip
Let us now outline the proof of the above expressions.
The proof will be based on a series of results which we have obtained
along the previous sections and which we now put together in their
generalized version:
\par\bigskip
{\sl 1. Permutation invariance}
\par\medskip
For any element $V$ of the permutation group of $N$
elements, $S_N$,
\begin{equation}
\label{vrhon}
\left[V, \rho^{\otimes N}\right]= 0,\qquad \forall V \epsilon S_N \ .
\end{equation}

\par\bigskip
{\sl 2. Spin invariance}
\par\medskip
With the following notation for the composite Hilbert
space,
\begin{equation}
\label{hilberts}
{\cal H}^{(N)} \equiv {\cal H}_A \otimes
{\cal  H}_B \otimes ... {\cal H}_N,
\end{equation}
for the corresponding local spin operators,
\begin{eqnarray}
\label{sabn}
\nonumber
\vec S_A &\equiv& {1\over 2}\vec {\sigma} \otimes I^{\otimes N-1},\\
\nonumber
\vec S_B &\equiv& {1\over 2} I \otimes \vec {\sigma} \otimes
I^{\otimes N-2},\\
\vec S_N
 &\equiv& {1\over 2} I^{\otimes N-1} \otimes \vec {\sigma}, 
\end{eqnarray}
and for the partial and total spin operators
\begin{equation}
\label{sms}
\vec {S}_{(M)} \equiv \sum^M_{x=A} \vec {S}_x,\quad A<\forall
M < N \qquad,\qquad
\vec {S} \equiv \vec {S}_{(N)} 
\end{equation}
the spin invariances read
\begin{equation}
\label{comm}
\left[\vec {S^2}, \rho^{\otimes N}\right]=
\left[\vec {S}^2_{(M)}, \rho^{\otimes N}\right]=
\left[\vec{S}^2_A, \rho^{\otimes N}\right]=0\ .
\end{equation}
They are an immediate consequence of the following relatively
straightforward result,
\begin{equation}
\label{rhon}
\rho(\vec {b})^{\otimes N}= p_N 
\left(\vec {S}\cdot \vec {b}\right)\ ,
\end{equation}
where $p_N (x)$ is a polynominal of degree $N$ in $x$.
\par\bigskip
{\sl 3. Direct sum decomposition}
\par\medskip
Since
\begin{equation}
\label{directsum}
\left[\vec {S}^2, \vec {S}^2_{(M)}\right]=
\left[\vec {S}^2_{(M)}, \vec
{S}^2_{(L)}\right]=0\qquad \forall M, L 
\end{equation}
the total Hilbert space can be written as a direct sum
\begin{equation}
\label{totalh}
{\cal H}^{(N)}= \oplus_{s, \{ s_{(M)}\} } E_{s, \{ s_{(M)}\} } 
\end{equation}
where $E_{s, \{ s_{(M)}\} }$ are the eigenspaces of $\vec {S}^2$ and $\vec
{S}^2_{(M)}, N>\forall M> A$, with eigenvalues $s(s+1), \{ s_M
(s_M +1)\}$ ordered with decreasing $M$, respectively. For instance, for $N = 4$,
\begin{eqnarray}
{\cal H}^{(N=4)}= E_{2, {3\over 2}, 1}\oplus \qquad  \qquad (s=2) \nonumber \\
 E_{1, {3\over 2}, 1}\oplus E_{1, {1\over 2}, 1}\oplus E_{1, {1\over 2}, 0}\oplus \qquad \qquad (s=1) \nonumber \\
E_{0, {1\over 2}, 1}\oplus E_{0, {1\over 2}, 0} \qquad \qquad \,\,\,\,\,\,  (s=0).
\end{eqnarray}
Of course only
those eigenvalues consistent with the spin composition rules appear.
\par\bigskip
{\sl 4. Permutation group equivalence}
\par\medskip
For a given $s<{N\over 2}$ all the spaces $E_{s,
\{s_{(M)} \} }$ corresponding to it can be obtained from one of them with
the help of the elements of the permutation group. The one which we
 retain for our proof as reference space is the one with the maximal number of
vanishing  partial spins,
\begin{equation}
\label{ees}
E_{s, s-{1\over 2}, s-1, ...0, {1\over 2}, 0} \qquad
({\rm with}\ {N\over 2}-s\ \ {\rm zeros}).
\end{equation}
There are as many of these equivalent spaces as the dimension of the
irreducible representation of $S_N$ in a space of total spin $s$.
\begin{equation}
\label{spaces}
d_N (s)= \left({N\atop {N\over 2} + s}\right) {2s + 1\over {N\over 2} + s+ 1}\ .
\end{equation}
One can check the dimensional consistency of the previous expression
from eq. (\ref{totalh}),
\begin{equation}
\label{checkdim}
2^N=\sum^{{N\over 2}}_{s=s_0} (2s+1) \ d_N (s) \qquad s_0=0\ {\rm or}\ 
{1\over 2}
\end{equation}
\par\bigskip

{\sl 5. Optimal pure state measuring strategy}
\par\medskip
In each of the reference spaces of the type of eq. (\ref{ees})
where any vector is of the form
\begin{equation}
\label{vecform}
\vert\sigma\rangle^{\otimes{N\over 2}-s}\otimes \vert \psi\rangle\qquad,
\qquad \vert\psi\rangle \in C^{2\otimes 2s}
\end{equation}
the best measuring strategy turns out to be the one
 corresponding to $2s$ copies of
an unknown pure state [1, 2, 3], and thus projects onto states of the form
\begin{equation}
\label{stateform}  
\vert \sigma \rangle^{\otimes {N\over 2} - S}\otimes \vert \hat
{n}\rangle^{\otimes 2 s} 
\end{equation}
Notice that the singlets act as an identity in the reference space of
eq. (\ref{ees}) and that the states (\ref{stateform}) are the ones
in eq. (\ref{vecform}) with less entanglement. 
 From here, and recalling eq. (\ref{spaces}), one readily
obtains eqs. (\ref{ogen}) and (\ref{idengen}). The fact that the guesses of eq.
(\ref{guessgen}) can be grouped together due to the permutation equivalence, and
thus have to be made only for the reference space, has been taken into
account already in writing eq. (\ref{ogen}). Notice that the operators of eq.
(\ref{ogen}) are of rank $d_N(s)$.

\bigskip
We are now ready to perform the final computation of 
\begin{equation}
\label{fmaxgenagain}
\overline {F}^{(N)}_{max} = \sum^{{N\over 2}}_{s=s_o}
\sum^{n_{ps}^{(2s)}}
_{i=1} \int d\Omega \int^1_0 db\ b^2 f(b) {\rm Tr}\left(
O_{N, s, i} \rho^{\otimes N}\right)
F (\rho, \rho_{N, s, i}).
\end{equation}
From
\begin{equation}
\label{tracegen}
{\rm Tr}\left(O_{N, s, i} \rho^{\otimes N}\right)
 = c^2_{s, i} d_N (s) \left({1- b^2\over
4}\right)^{{N\over 2}- s} \left({1+ \vec {b} \cdot \hat {n}_{s, i}\over 2}
\right)^{2s},
\end{equation}
which is obtained from eq. (\ref{probabi}), (\ref{ogen}) and 
(\ref{spaces}), eq. (\ref{fmaxgenagain})
can be written as
\begin{eqnarray}
\nonumber
\overline {F}^{(N)}_{max}&=& \sum^{N\over 2}_{s=s_o} (2s+1) d_N (s) \int
d\Omega \int^1_0 db \ b^2 f(b) \left({1-b^2\over 4}\right)^{{N\over 2}- s} 
\left({1+\vec
{b}\cdot \hat {n} \over 2}\right)^{2s}\\
\label{final}
&&{1\over 2} \left(1+ r_{N, s}\ \vec {b}\cdot \hat {n} + \sqrt {1-b^2}
\sqrt {1- r^2_{N, s}}\right)
\end{eqnarray}
where we have used eq. (\ref{condgen}), as the contributions corresponding to
different $\it {i}$ are the same, eq. (\ref{rhorhoi}) for the fidelity, and
where the subindices of $\hat {n}_{s,i}$ have been dropped, given their
irrelevance at this stage of the computation. In eq. (\ref{final}) the first
term gives $1\over 2$ and the other two depend on $r_{N,s}$, which is
fixed by maximization. Choosing $\hat {n}$ in the direction of the
z-axis, and with the definitions of eq. (\ref{ges}), one immediately obtains
eq. (\ref{rngen}) and finally our main result eq. (\ref{fmaxgen}).
 The result refering
to the number of outputs of minimal measurements, eq. (\ref{nmingen}), follows
from our point 5 above.

\section{Conclusions}
\renewcommand{\theequation}{6.\arabic{equation}}
\setcounter{equation}{0}

We have built the optimal and minimal measuring
strategy for $N$ copies of an unknown mixed state prepared according to
a known, isotropic, but otherwise arbitrary probability distribution.
The strategy is universal, i.e. independent of the probability
distribution. Except for one single copy, optimal measurements have to
be generalized measurements. We have obtained a closed expression
for the maximal averaged mean fidelity, and the associated
minimal number of outcomes. In obtaining these expressions some
interesting windfall results emerged. They are:

1) Best guesses are not universal. They are pure states only if the unknown
state is known to be pure. 

2) Optimal measurements require projecting onto total spin
eigenspaces, and within each such subspace, onto total spin eigenstates with maximal total spin component in some direction. This allows to relate them with
optimal measurements corresponding to a smaller number of copies of
unknown pure states.

3) Optimal measurements which are minimal have,
beyond two copies, outcomes associated with positive operators of rank 
larger than one and, beyond
three copies, less outcomes than dimensions of the Hilbert space. These
optimal measurements are thus incomplete! Completing them is
useless.

Our results also set the limits to optimal cloning of
mixed states. The techniques developed here for dealing with copies of mixed states will be useful for solving related problems.

After finishing this work we learned from Ignacio Cirac that he has done, together with Artur Ekert and Chiara Macchiavello, somewhat similar work using basically the same techniques.

\bigskip
\centerline{\bf Acknowledgments}

This work was started during the Benasque Center for Physics '98 session. Financial support from CIRYT, contract AEN98-0431 and CIRIT, contract 1998SGR-00026 is acknowledged. G.V. acknowledges a CIRIT grant 1997FI-00068PG. R.T. thanks Ignacio Cirac and Peter Zoller for hospitality in Innsbruck.


\begin{thebibliography}{99}
\bibitem{MP}
S. Massar and S. Popescu, Phys. Rev. Lett. {\bf 74} (1995) 1259.
\bibitem{DE}
R. Derka, V. Buzek and A. K. Ekert, Phys. Rev. Lett. {\bf 
80} (1998) 1571.
\bibitem{LPT}
J. I. Latorre, P. Pascual and R. Tarrach, Phys. Rev. Lett.
{\bf 81}(1998) 1351.
\bibitem{PERES}
A. Peres ``Quantum theory, concepts and methods'', Kluwer Acad.
Publ. 1995.
\bibitem{JOZSA}
R. Jozsa, Jour. Mod. Opt. {\bf 41} (1994) 2315.
\bibitem{DAVIES}
E.B. Davies, IEEE Trans. Inform. Theory IT-24 (1978) 596. 
\bibitem{HB}
M. Hillary and V. Buzek, Phys. Rev. {\bf A56}
(1997)  1212.
\bibitem{BDiV}
D. Bruss, D.P. DiVincenzo, A. Ekert, Ch.A. Fuchs, Ch. Macchiavello
and J. Smolin,
Phys. Rev. {\bf A57} 2368 (1998).
\bibitem{GISIN}
N. Gisin, Phys. Lett.  {\bf A242}(1998)  1.
\bibitem{GM}
N. Gisin and S. Massar, Phys. Rev. Lett. {\bf 79} 
(1997) 2153.
\bibitem{BCFJS}
H. Barnum, C.M. Caves, Ch.A. Fuchs, R. Jozsa and B. Schumacher,
Phys. Rev. Lett. {\bf 76} (1996) 2818.
\bibitem{strong}
Strong symmetry means $V\rho^{(2)}_c V=\rho^{(2)}_c$ which
implies $t_{ij}=t_{ij}$ and strong isotropy
means that $t_{ij}$ is a tensor which only depends on the vector 
$\vec {b}$ and the invariant
tensors. It is obvious that optimal cloners can be chosen to be symmetric
and that as $f(b)$
is isotropic optimal cloners will not break this
isotropy.
\bibitem{WERNER}
R.F. Werner, Phys. Rev. {\bf A58} (1998)  1827.

\end{thebibliography}
\end{document}